\documentclass[twocolumn,trackchanges]{aastex63}
\usepackage{amsmath}
\usepackage{array}
\usepackage{booktabs}
\usepackage{lipsum}
\usepackage{textgreek}
\usepackage[figuresright]{rotating}
\usepackage{hyperref}
\usepackage{mathrsfs}
\usepackage{natbib}
\usepackage{soul}

\shortauthors{Huang \& Li}
\graphicspath{{./}{figures/}}

\begin{document}

\title{Constraints on Hadronic Contribution to LHAASO Sources with Neutrino Observations}

\author[0000-0001-8510-2513]{Tian-Qi Huang}
\affiliation{Department of Astronomy, School of Physics, Peking University, Beijing 100871, China}

\author{Zhuo Li}
\affiliation{Department of Astronomy, School of Physics, Peking University, Beijing 100871, China}
\affiliation{Kavli Institute for Astronomy and Astrophysics, Peking University, Beijing 100871, China}



\begin{abstract}
LHAASO detected 12 gamma-ray sources above 100 TeV which are the possible origins of Galactic
cosmic-rays. We summarize the neutrino measurements by IceCube and ANTARES in the vicinity of LHAASO sources to constrain the contribution of hadronic gamma-rays in these sources. We find that the current observations constrain that the hadronic gamma-rays contribute no more than $\sim60\%$ of the gamma-rays from Crab Nebula. Gamma-rays from two LHAASO sources, LHAASO J1825-1326 and LHAASO J1907+0626, are dominated by leptonic components up to $\sim200$ TeV, under the hypotheses in the analysis by IceCube. The uncertainties of the constraint on the hadronic gamma-ray emission are discussed. We also constrain the total 100 TeV gamma-ray emission from TeV PWNe relying on the remarkable sensitivity of LHAASO at that energies. 
\end{abstract}







\section{Introduction} \label{sec:intro}


Cosmic-rays extending to several PeV energies are believed to originate from Galactic sources, called PeVatrons. It is hard to localize PeVatrons with only cosmic-ray observations, because charged cosmic-rays below several PeV are deflected frequently by magnetic field while traveling in the Galaxy. Fortunately, PeV cosmic-rays produce high energy ($\rm >$100 TeV) gamma-rays and neutrinos in the collisions with baryon background through hadronuclear ($pp$) interactions or radiation background through photohadronic ($p\gamma$) interactions. Those gamma-rays and neutrinos are good tracers of PeVatrons, as both of them travel in a straight line after escaping from the sources.

Gamma-rays can be produced through hadronic process ($pp$ or $p\gamma$) and/or leptonic process (inverse Compton scattering). The production mechanisms of observed gamma-rays usually are unclear. However, compared with TeV gamma-rays, the production of $\ga100$TeV gamma-rays with inverse Compton scatterings suffers more stringent Klein–Nishina suppression.  The sources with $\sim100$ TeV emission have more chance to be cosmic ray PeVatrons \citep[e.g.][]{2016Natur.531..476H, 2020ApJ...896L..29A, 2021NatAs.tmp...41T}.

 
 







The Large High Altitude Air Shower Observatory (LHAASO) is a hybrid extensive air shower array for cosmic-ray and gamma-ray studies. Relying on the unprecedented sensitivity to gamma-rays around hundreds TeV, the Kilometer Square Array (KM2A) of LHAASO detected 12 gamma-ray sources above 100 TeV \citep{2021Natur.594...33C}. They also observed PeV gamma-rays from the direction of Cygnus region and Crab nebula \citep{425}. These LHAASO gamma-ray sources provide us a group of target sources to find out the cosmic ray PeVatrons. We should look for neutrino signals from these sources to pin down the answer.

The IceCube neutrino observatory is a cubic-kilometer array in the deep ice at the South Pole and currently the most
sensitive detector for neutrinos from TeV to PeV. IceCube has carried out some approaches to search for neutrino sources from known Galactic sources, but only gives upper limits to the flux. First, in the \textit{Source List Search} (SLS), they select some known astrophysical sources to be the \textit{Neutrino Source Candidates} (NSCs), and searches for neutrino signals from them individually \citep[e.g.][]{2019EPJC...79..234A}. The signals are not significant enough. Second, since stacking may help to enhance signal to noise ratio and discover weak signals, IceCube also tries to search for neutrinos over some catalogs of known sources, i.e., \textit{Stacked Source Search} (SSS) \citep[e.g.][]{2020PhRvL.124e1103A}. Still, only upper limits are given to the neutrino flux so far. However, as neutrinos are produced together with gamma-rays in the hadronic interactions, one can derive the upper limits of neutrino flux by that of hadronic gamma-ray flux, and hence constrain the contribution of hadronic process.



With the goal of identifying the Galactic cosmic ray Pevatrons from the LHAASO detected 100 TeV sources, in this work we use the neutrino measurements of NSCs in the vicinity of LHAASO sources, and constrain the hadronic contributions to gamma-ray emissions. In \autoref{sec:source}, we introduce the neutrino source searches used in this paper and the criteria to select the NSCs. In \autoref{sec:method}, we show the method to derive the upper limits on hadronic gamma-ray flux. In \autoref{sec:results}, we compare the gamma-ray observations with upper limits of hadronic gamma-ray flux. In \autoref{sec:discussion}, we discuss previous works and evaluate the uncertainties in the analyses. We finally give conclusions in \autoref{sec:conclusions}.



\section{Source Selection}\label{sec:source}

IceCube uses track-like neutrino events to measure the astrophysical neutrino flux from NSCs. The median location errors of these neutrinos range from 0.25 degree (above PeV) to 1 degree (sub-TeV) \citep{2020PhRvL.124e1103A}, while most LHAASO sources represent diffuse structures with angular extensions as large as 1 degree except for the Crab Nebula and LHAASO J2108+5157 \citep{2021Natur.594...33C}. Astrophysical neutrinos from the directions of LHAASO sources probably have the same origins with those of gamma-rays.


We select the NSCs within 1 degree from the measured center of a LHAASO source (see \autoref{table01}).  The NSCs are taken from three SLSes: the ten-year search \citep{2020PhRvL.124e1103A}, the eight-year search \citep{2019EPJC...79..234A} and the combined search by ANTARES \& IceCube \citep{2020ApJ...892...92A}. The upper limits of neutrino flux given by SLSes depend on the hypotheses on spectral shapes (usually assumed as a single power law) and point-source hypothesis. We also choose the NSCs from two SSSes: the TeV Pulsar Wind Nebulae (PWNe) search \citep{2020ApJ...898..117A} and the Galactic catalog searches \citep{2017ApJ...849...67A} including five catalogs (one Milagro catalog, one HAWC catalog and three Supernova Remnant (SNR) catalogs). The upper limits given by SSSes depend on not only the spectral shapes and assumed source extensions, but also the weighting factors of neutrino flux from individual sources in the certain catalog. The weighting factor is the fraction of neutrino flux from individual source and depends on the weighting scheme used. The details about the weighting schemes are explained in \autoref{appendix:A}.

Nine LHAASO sources have NSCs in the neighborhood. In order to compare the gamma-ray emission with neutrino emission in the range of TeV to PeV, we utilize previous TeV gamma-ray measurements of the certain NSCs to obtain the energy spectra at lower energies. The details about the TeV gamma-ray observations of the NSCs are described in \autoref{appendix:B}.


\section{Method}\label{sec:method}


In hadronuclear scenarios, the differential flux of gamma-rays and neutrinos ($\Phi_{\gamma,\,\nu}=dN_{\gamma,\,\nu}/dE_{\gamma,\,\nu}$) are related as
\begin{equation}
\frac{1}{3}\sum_{\rm \alpha} \frac{dN_{\nu_{\rm \alpha}+\bar{\nu}_{\rm\alpha}}}{dE_{\rm \nu}}=K_{\rm \pi}\frac{dN_{\rm \gamma}}{dE_{\rm \gamma}} \label{relationship}
\end{equation}
where $\rm \alpha$ represents the neutrino flavor, $E_{\rm \gamma}=2E_{\rm \nu}$ and $K_{\rm \pi}\simeq2$ is the number ratio of charged to neutral pions. For an $dN/dE\propto E^{-\gamma}$ spectrum, the upper limit on hadronic gamma-ray flux at 100 TeV can be expressed as
\begin{equation}
\Phi_{\rm \gamma}^{\rm UL}(100~{\rm TeV})=\frac{1}{2}\Phi_{\rm \nu_\mu+\bar{\nu}_\mu}^{\rm UL}(E_{\rm \nu})\left(\frac{50~{\rm TeV}}{E_{\rm \nu}}\right)^{-\gamma}
\label{eqm02}
\end{equation}
where $\Phi_{\rm \nu_\mu+\bar{\nu}_\mu}^{\rm UL}$ is the 90\% upper limit of neutrino flux. 

Gamma-rays are partially absorbed in the propagation through interstellar radiation field (ISRF) and cosmic microwave background (CMB) due to pair productions ($\gamma\gamma\rightarrow e^{+}e^{-}$). Such absorption is considered in the conversion of neutrino upper limit into gamma-ray upper limit, if the distance to the NSC is available. The ISRF energy density is taken from \citet{2017MNRAS.470.2539P}. Gamma-ray absorption around the source is not considered.

\section{Results}\label{sec:results}

\begin{figure*}[htbp!]
\begin{center}
\includegraphics[width=0.78\textwidth]{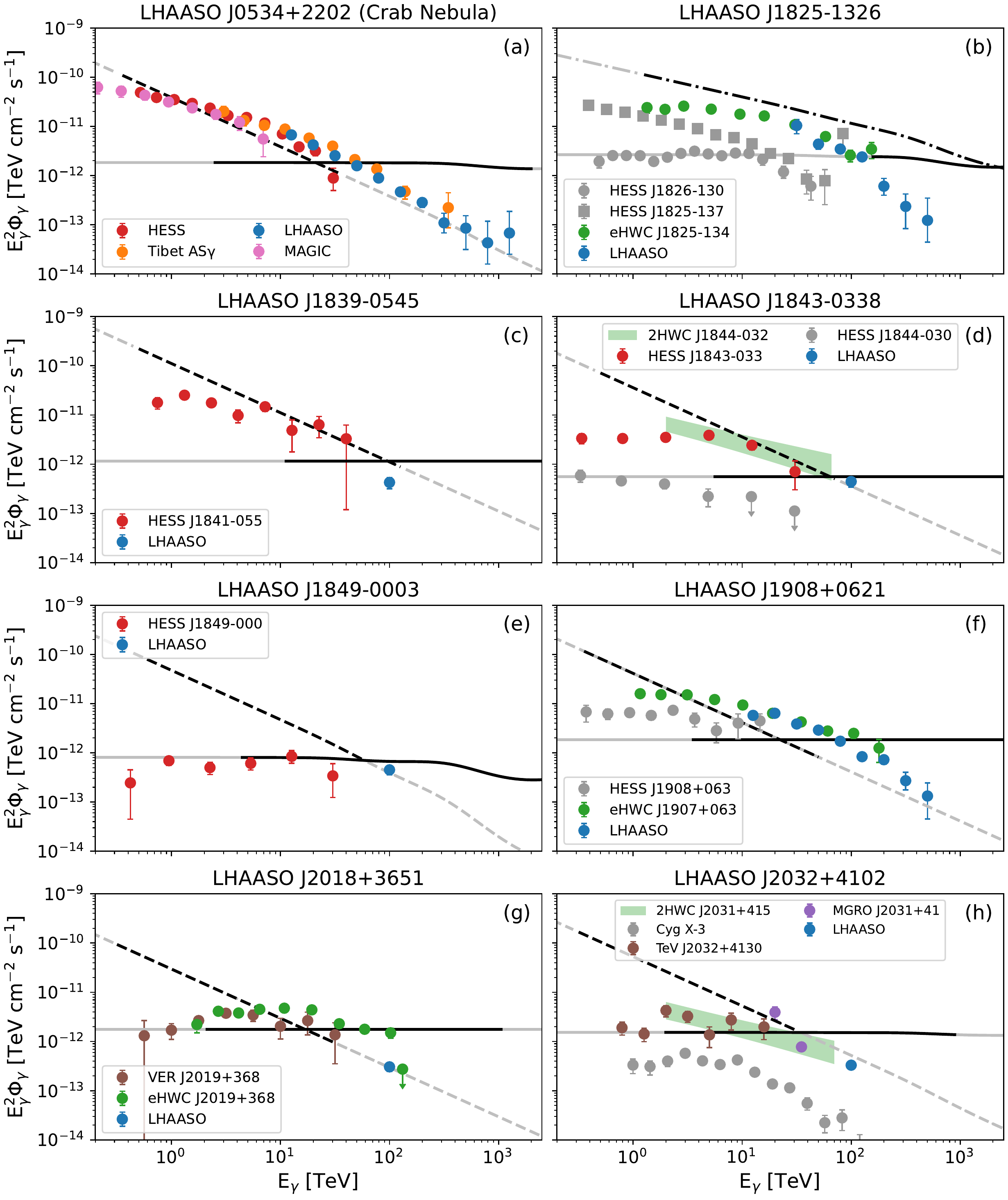}
\caption{ Comparison between gamma-ray flux by LHAASO observations and the upper limits of hadronic gamma-ray flux derived from SLSes. The grey lines present the 90\% upper limits of hadronic gamma-ray flux, where the dark parts mark the central 90\% energy ranges according to neutrino observations. The solid, dot-dashed and dashed lines correspond to $E^{-2.0}$, $E^{-2.5}$ and $E^{-3.0}$ neutrino spectra, respectively. Blue symbols are LHAASO observations \citep{2021Natur.594...33C,425}. Other symbols and shaded areas are the gamma-ray measurements of (a) Crab Nebula by Tibet AS$\gamma$ \citep{2019PhRvL.123e1101A}, H.E.S.S. \citep{2006A&A...457..899A} and MAGIC \citep{2008ApJ...674.1037A}; (b) HESS J1826-130 \citep{2020A&A...644A.112H}, HESS J1825-137 \citep{2018yCat..36210116H}, eHWC J1825-134 \citep{2020PhRvL.124b1102A}; (c) HESS J1841-055 \citep{2008A&A...477..353A}; (d) HESS J1843-033 and HESS J1844-030 \citep{2018A&A...612A...1H}, 2HWC J1844-032 \citep{2017ApJ...843...40A}; (e) HESS J1849-000 \citep{2018A&A...612A...1H}; (f) HESS J1908+063 \citep{2009A&A...499..723A}, eHWC J1907+063 \citep{2020PhRvL.124b1102A}; (g) VER J2019+368 \citep{2018ApJ...861..134A}, eHWC J2019+368 \citep{2020PhRvL.124b1102A}; (h) Cyg X-3 \citep{2020JPhCS1468a2092S}, TeV J2032+4130 \citep{2014ApJ...783...16A}, MGRO J2031+41 \citep{2007ApJ...664L..91A, 2009ApJ...700L.127A}, 2HWC J2031+415 \citep{2017ApJ...843...40A}. The gamma-ray absorption due to ISRF and CMB is considered if the distance is available. }\label{figure01}
\end{center}
\end{figure*}

\begin{figure*}[htbp!]
\begin{center}
\includegraphics[width=0.98\textwidth]{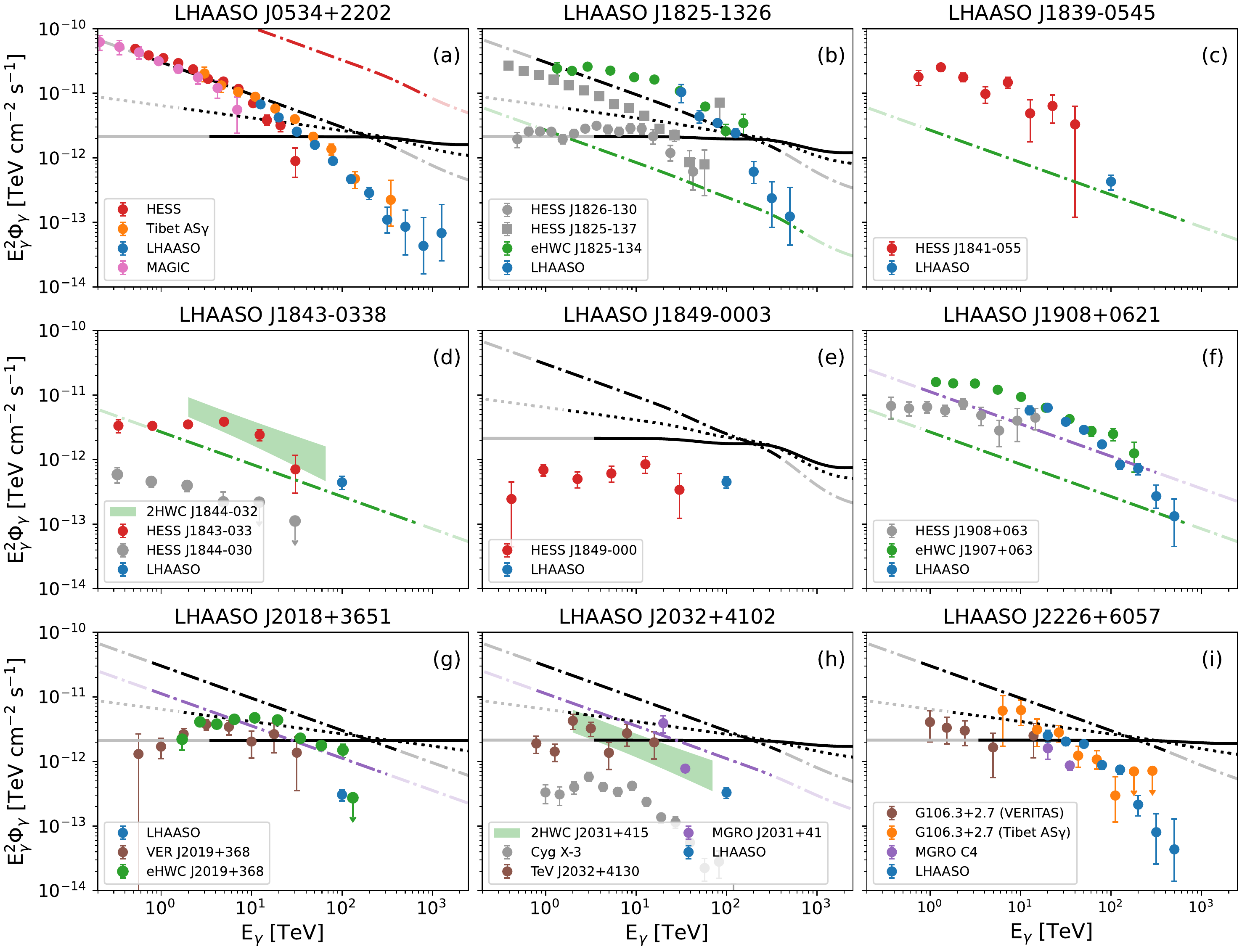}
\caption{Comparison between gamma-ray flux by LHAASO observations and the upper limits of hadronic gamma-ray flux derived from SSSes. Lines show the 90\% upper limits of hadronic gamma-ray flux, where the dark parts mark the central 90\% energy ranges according to neutrino observations. The solid, dotted, dot-dashed and dashed lines correspond to $E^{-2.0}$, $E^{-2.19}$, $E^{-2.5}$ and $E^{-3.0}$ neutrino spectra respectively. The colors of lines represent different source catalogs: black lines correspond to the TeV PWNe catalog (inverse age weighting); red lines to the catalog of SNRs with associated PWN; green lines to the HAWC catalog and purple lines to the Milagro catalog. The symbols in panel (i) are gamma-ray flux from G106+2.7 observed by VERITAS \citep{2009ApJ...703L...6A} and Tibet $\rm AS\gamma$ \citep{2021NatAs.tmp...41T}, MRGO C4 \citep{2007ApJ...664L..91A, 2009ApJ...700L.127A} and LHAASO J2227+6057 \citep{2021Natur.594...33C}. Symbols and shaded areas in other panels are the same as those in \autoref{figure01}. }\label{figure_stacking}
\end{center}
\end{figure*}

\begin{figure*}[htbp!]
\begin{center}
\includegraphics[width=0.84\textwidth]{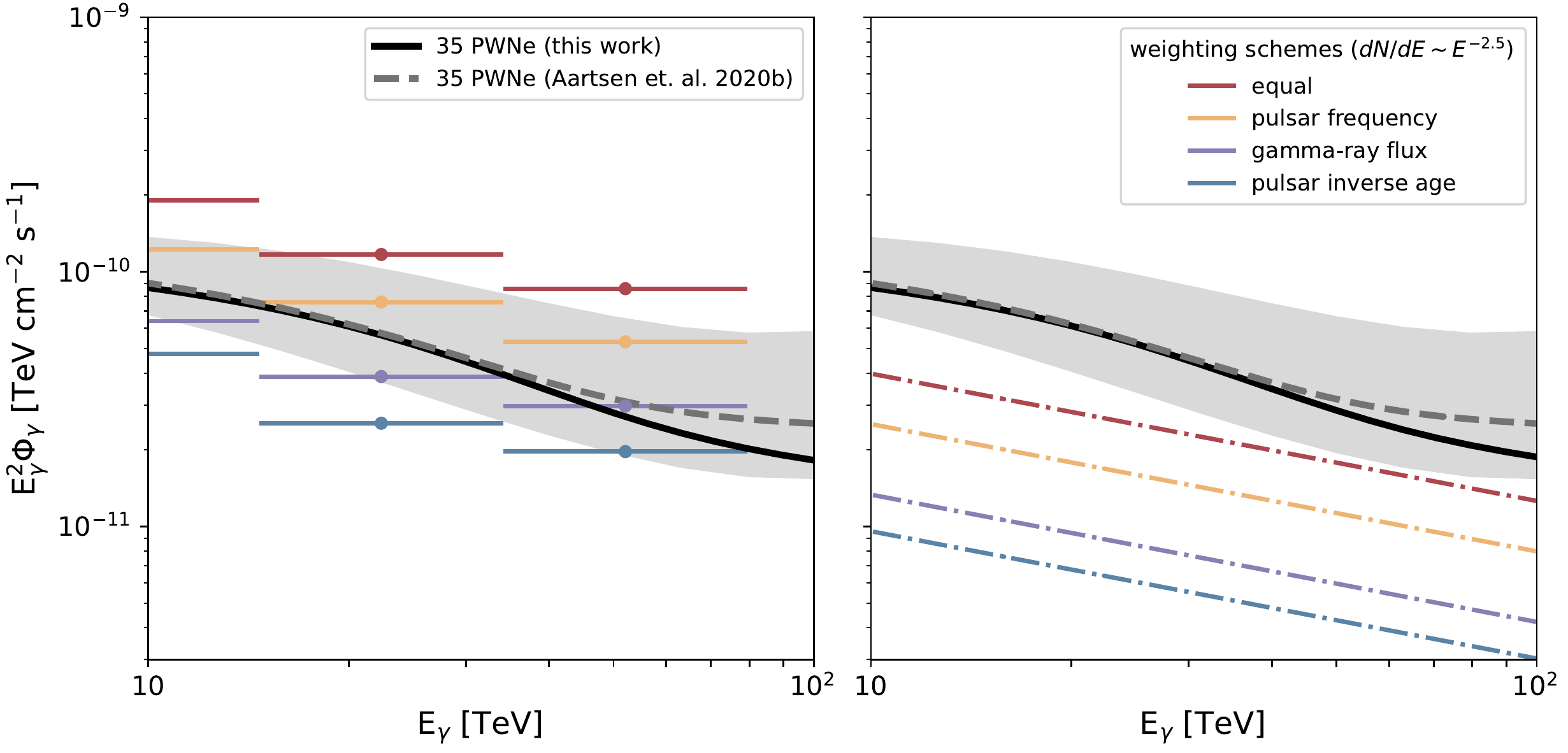}
\caption{The cumulative gamma-ray flux from TeV PWNe compared with the upper limits of hadronic gamma-ray flux. There are 35 PWNe in the catalog. The cumulative gamma-ray flux from all the PWNe (black solid line) is 26\% lower than the cumulative gamma-ray flux extrapolated from low  energy  before  LHAASO  measurement (grey dashed line). The light grey shaded area indicates the uncertainty of the extrapolated spectrum.  Colorful lines show the upper limits on the hadronic gamma-ray emission for four weighting schemes \citep{2020ApJ...898..117A}.}\label{figure_pwn}
\end{center}
\end{figure*}

The constraints on the hadronic gamma-ray emissions derived from SLSes are shown in \autoref{figure01}. If there are more than one NSCs associated with the LHAASO source, we only show the upper limit derived from the NSC closest to the center. The upper limit for $E^{-2}$ spectrum is generally consistent with or higher than the observed gamma-ray flux around 100 TeV in each panel. The upper limits for $E^{-3}$ and $E^{-2.5}$ are also generally consistent with the observed gamma-ray flux from 10 TeV to 100 TeV but much higher at the energies below 10 TeV (see panel b, c, d, f, and g). We also show in the figure the energy range corresponding to where 90\% of signal neutrino events will concentrate in. This energy range is determined by the assumed neutrino spectrum, as well as the effective area of the neutrino detector. The gamma-ray absorption due to ISRF and CMB is not significant around 100 TeV.

The constraints on the hadronic gamma-ray emissions derived from SSSes are shown in \autoref{figure_stacking}. The hadronic gamma-ray flux from individual source in the catalog should be lower than the upper limit derived from the stacking neutrino flux. The upper limit given by the HAWC catalog set strong constraints on LHAASO J1825-1326 (panel b) and LHAASO J1908+0621 (panel f) above 100 TeV. The leptonic component is even dominant around 200 TeV. For the TeV PWNe, the only shown are the upper limits for inverse age weighting, because they are lower than those for the other weighting schemes, and give stronger constraints on the hadronic gamma-rays \citep{2020ApJ...898..117A}. The constraints above are reliable when the hypotheses in the SSSes are consistent with the real case. In the HAWC catalog search, the sources are assumed to be $0.5^{\circ}$ extended and the neutrino flux following an $E^{-2.5}$ spectrum is assumed to be proportional to the gamma-ray flux at pivot energy for each source. The assumption on neutrino flux ratio brings large uncertainty to the upper limit estimate (see \autoref{sec:uncertain}). 

We further estimate the cumulative gamma-ray flux from TeV PWNe with LHAASO observations and compare it to the upper limits given by the TeV PWNe search in the \autoref{figure_pwn}. In the TeV PWNe search, \cite{2020ApJ...898..117A} compared the differential upper limits for different weightings with the cumulative gamma-ray flux extrapolated from TeV observations to constrain the hadronic component from TeV to 100 TeV. Six of these PWNe are associated with LHAASO sources, while seven of them are not detected but in the data taken region of LHAASO-KM2A (sky declination band $-15^{\circ}<\delta<75^{\circ}$). With the observations at 100 TeV, we correct the extrapolated gamma-ray spectrum as the black solid line in \autoref{figure_pwn} (details in \autoref{appendix:C}). The cumulative gamma-ray flux at 100 TeV is 26\% lower than the extrapolated value (grey dashed line) but still consistent with the uncertainty (grey shaded area).


\section{Discussion}\label{sec:discussion}

\subsection{Compare with previous works}

We compare the neutrino spectrum model of Crab Nebula in \cite{2007ApJ...656..870K} with the neutrino observation results in the ten-year search. First of all, the upper limit on neutrino flux (with a certain spectral shape) $\Phi^{90\%}_{\nu}$ is related to the upper limit on the signal event number $n_{90}$ as
\begin{equation}\label{eq03}
	n_{\rm 90}=\int dt\int dE_{\rm \nu} \Phi_{\rm \nu_\mu+\bar{\nu}_\mu}^{90\%}A_{\rm eff}(E_{\rm \nu},\delta_{s})
\end{equation}
where $A_{\rm eff}$ is the effective area for neutrinos from the direction of source ($E_\nu$ is the neutrino energy and $\delta_s$ is the decliantion of source). The effective area is available in the IceCube muon-track data from 2008 to 2018 \citep{2021arXiv210109836I}.

\begin{figure}[htbp!]
\begin{center}
\includegraphics[width=0.98\columnwidth]{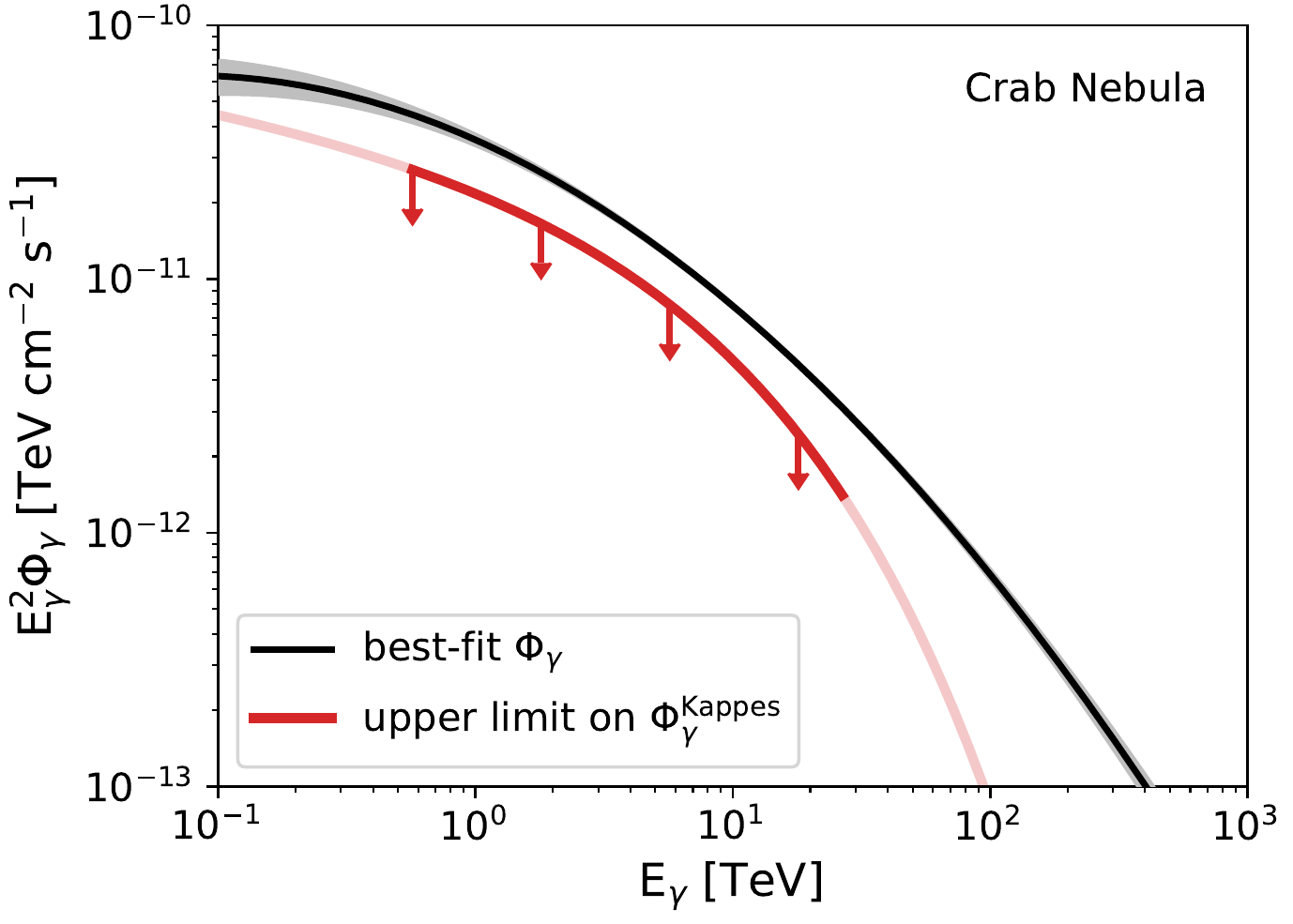}
\caption{ The best-fit gamma-ray spectrum $\Phi_{\gamma}$ (black line) of the Crab Nebula and the upper limit of hadronic gamma-ray flux $\Phi^{\rm Kappes}_{\rm \gamma}$ (red line). The dark part of red line shows the central 90\% energy range of signal events.}\label{figure04}
\end{center}
\end{figure}

In the seven-year search one has $n_{90}=7.0$ and $21.2$ for the spectra of $E^{-2}$ and $E^{-3}$, respectively \citep{2017ApJ...835..151A}. The two values both decrease in the ten-year search \citep[][$n_{90}=5.8$ and $20.4$, respectively]{2020PhRvL.124e1103A}.
Note that the local pre-trial p-value for Crab Nebula is larger in the ten-year search (from 0.34 in the seven-year search to 0.49), indicating more background like. Therefore, we assume that $n_{90}$ using Kappes model decreases between two searches. 
For the Kappes model a value $n_{90} = 20.5$ is derived from the seven-year limit \citep{2017ApJ...835..151A} with \autoref{eq03}, which can be considered as the upper limit for ten-year search, and hence results in a flux limit (red line in \autoref{figure04}) by \autoref{eq03} and \autoref{relationship}. 


We fit the gamma-ray observations with a log-parabola model $(E/10\,\rm{TeV})^{-\gamma_{1}-\gamma_{2}\,\rm{log}(E/10\,\rm{TeV})}$ (black line in \autoref{figure04}). The number of hadronic gamma-rays in the central 90\% energy range contributes $\rm 62^{+5}_{-4}\%$ to the total gamma-rays at most, while the energy of hadronic gamma-rays contributes $\rm 62{\pm}4\%$ at most. Thus, the current observations suggest that hadronic contribution can not account for the bulk of the gamma-ray emission from Crab Nebula.

Furthermore, there have been many discussion in the literatures on the hadronic gamma-ray and neutrino emission from potential Galactic PeVatrons and the Galactic plane, say, for PWNe \citep[e.g.][]{2003A&A...402..827A}, for SNRs \citep[e.g.][]{2015APh....65...80M}, for young massive star clusters (YMCs) \citep[e.g.][]{2007PhRvD..75f3001A}, and for diffuse Galactic gamma-ray and neutrino emission \citep{2014PhRvD..90b3010A, 2014PhRvD..89j3002N}. Some TeV gamma-ray sources have been well studied to evaluate the probability of being PeVatrons, like the Cygnus Region \citep[e.g.][]{2007PhRvD..75f3001A}, MGRO J2019+37 \citep[e.g.][]{2007PhRvD..75h3001B} and MGRO J1908+06 \citep[e.g.][]{2017APh....86...46H}. However, the LHAASO detected $100$ TeV sources provide a group of sample sources for candidate PeVatrons. The high sensitivity of LHAASO at 100 TeV ranges also help to put stringent constraint on the PeVatron models. A wide range of parameter space in the Crab Nebula model by \cite{2003A&A...402..827A} can be constrained by LHAASO observation at 100 TeV.

\subsection{Uncertainties in analysis}\label{sec:uncertain}

The upper limit on the hadronic gamma-ray flux depends on the hypotheses of source extension. As the signals are not that significant over the background, the upper limit ratio for Extended-Source (ES) hypothesis approximates as $\Phi^{90\%}_{\rm PS}\times(\Omega_{\rm ES}/\Omega_{\rm PSF})^{1/2}$, where $\Phi^{90\%}_{\rm PS}$ is the upper limit for Point-Source (PS) hypothesis, $\Omega_{\rm ES}$ is the angular size of extended source and $\Omega_{\rm PSF}$ is the angular size of point spread function. The angular size is defined as $\Omega_{\rm ES}=\pi(\sigma_{s}^{2}+\sigma_{\rm IC}^{2})$ and $\Omega_{\rm PSF}=\pi\sigma_{\rm IC}^{2}$, where $\sigma_s$ is the source extension and $\sigma_{\rm IC}=0.64^{\circ}$ is given by \cite{2017APh....86...46H}. Three LHAASO sources (LHAASO J1825-1326, LHAASO J1908+0621  and LHAASO J2226+6057) have extension measurements with Gaussian template. They are $0.30^{\circ}$, $0.58^{\circ}$ and $0.36^{\circ}$ respectively \citep{2021Natur.594...33C}.  If their neutrino counterparts have the same extensions, their upper limits will be 10\%, 35\% and 15\% higher in comparison with the PS hypothesis. The upper limits given by SLSes will be harder to constrain the hadronic component for extended sources (panel b-h of \autoref{figure01}), while source extensions have been considered in the SSSes (e.g., HAWC sources in the Galactic catalog searches are assumed to be $0.5^{\circ}$ extended). 

The upper limit on the hadronic gamma-ray flux also relies  on  the  hypotheses of spectral shapes.  A Bayesian analysis of the IceCube muon-track data from 2008 to 2018 is carried out to test the impact of spectral shapes on the upper limits \citep{Huang2021b}. The hadronic component is assumed to follow the best-fit gamma-ray spectrum with a scaling factor. Five LHAASO sources are tested: LHAASO J0534+2202 (Crab Nebula), LHAASO J1825-1326, LHAASO J1908+0621, LHAASO J2018+3651 and LHAASO J2226+6057. Only the hadronic gamma-ray component of Crab Nebula is well constrained to be no more than 70\% of total gamma-ray flux. As for other sources, the upper limits on hadronic gamma-ray flux are higher than the flux observed.

The weighting schemes used in the SSSes influence strongly the upper limit estimate.  In the TeV PWNe search, the upper limit for the equal weighting is around 3 times of the one for the gamma-ray flux weighting (see the right panel of \autoref{figure_pwn}). In the stacking search for neutrinos from AGN and starburst galaxies, the upper limits for the equal weighting vary from 0.9 to 6.1 times of the upper limits for the gamma-ray flux weighting when changing source samples. In the HAWC catalog search, the neutrino flux from each source is assumed to be proportional to the gamma-ray flux at the pivot energy.  If the upper limits derived from the HAWC catalog search (green lines in \autoref{figure_stacking}) increase by a factor of six, we cannot simply use the upper limit on the stacking neutrino flux to constrain the hadronic gamma-ray flux from LHAASO J1908+0621 and LHAASO J1825-1326.

\cite{2007ApJ...656..870K} approximate the neutrino and gamma-ray relation as $\Phi_{\nu}(E)=(0.694-0.16\gamma)\Phi_{\gamma}(E)$, especially for primary proton spectrum being $E^{-\gamma}$ ($1.8<\gamma<3.0$) with a high energy exponential cutoff. If using this approximation, our upper limits on hadronic gamma-ray flux will be increased by 34\% (17\%) for the $E^{-2}$ ($E^{-3}$) spectrum.

The radiation background around LHAASO sources also influence the constraints on hadronic gamma-ray component. The gamma-ray sources associated with young massive stars may have a denser infrared background compared with ISRF. The higher optical depths around 100 TeV induce the lower upper limits on the hadronic gamma-ray flux observed. In photon-rich environment, $p\gamma$ interactions should also be considered in addition to the $pp$ interactions.

\section{Conclusions}\label{sec:conclusions}

We adopt the neutrino observation results in the directions within one degree from LHAASO sources, and transfer the neutrino upper limits into those of hadronic gamma-ray flux. The upper limits derived from the SLSes are marginally consistent with the gamma-ray flux for most LHAASO sources and cannot constrain the hadronic contributions stringently. The upper limits derived from the SSSes set strong constraints on the hadronic components above 100 TeV. These constraints depend on hypotheses of not only the spectral shapes and source extensions but also the weighting schemes. 

The main conclusions of this paper are summarized below:

\textbf{i)} The hadronic gamma-ray component contributes no more than 62\% for the Crab Nebula.

\textbf{ii)} LHAASO J1825-1326 and LHAASO J1908+0621 are leptonic dominant up to 200 TeV, following the assumption that each source in the HAWC catalog has an intrinsic extension $\la0.5$ degree and follows an $E^{-2.5}$ spectrum with neutrino flux proportional to its gamma-ray flux at pivot energy.

\textbf{iii)} The cumulative gamma-ray flux from TeV PWNe is lower  by 26\% at 100 TeV than the extrapolation from low energy before LHAASO measurement, leaving larger room for hadronic dominated emission.



The constraints on hadronic gamma-ray flux will be stronger as the neutrino statistics increase in the following 10 years. For IceCube, the time evolution of the $\rm5\,\sigma$ discovery potential is close to $1/T$ ($T$ is the observation time) \citep{2017ApJ...835..151A}. We take the discovery potential in the ten-year point source search as the baseline, corresponding to the data from 2008 to 2018. IceCube will be able to discover the point-like neutrino source two times fainter by 2028. If the intrinsic gamma-ray spectrum of LHAASO J1849-0003 follows $E^{-2}$ up to PeV, the neutrino observations with sensitivity two times better will offer more clues to answer the origins of these high energy gamma-rays.

A stacking search for astrophysical neutrinos from the twelve LHAASO sources is required, which will increase the significance of signals and constrain the hadronic component more strictly. As for the LHAASO sources in the southern sky ($\delta<-5^{\circ}$), the joint search of IceCube and ANTARES will improve the sensitivity by a factor ${\sim2}$ compared to both individual analyses \citep{2020ApJ...892...92A}. The instrumented volume of high energy neutrino detectors will be three times larger when KM3NeT/ARCA (two blocks) \citep{2016JPhG...43h4001A} and Baikal-GVD (15 clusters) \citep{2020arXiv201203373S} fully operate in the next 5 to 10 years.

\section*{Acknowlegdments}
We thank Qinrui Liu, Jon Dumm and Ruo-Yu Liu for useful discussions. This work is supported by the Natural Science Foundation of China (No. 11773003, U1931201).

\bibliography{sample63}{}
\bibliographystyle{aasjournal}

\begin{sidewaystable}[htbp!]
\begin{center}
\begin{tabular*}{\textwidth}{l l c c c l l l l l}
\hline
\hline
LHAASO source     &Neutrino Source  & Distance  & Angular Distance & Extension  & $\Phi_{0,\gamma=2.0}^{90\%}$ & $\Phi_{0,\gamma=2.19}^{90\%}$ & $\Phi_{0,\gamma=2.50}^{90\%}$ & $\Phi_{0,\gamma=3.0}^{90\%}$ & Neutrino Source Search\\
        &Candidate        & [kpc] & [deg] & [deg]        &\multicolumn{4}{c}{$\rm [10^{-17}\,TeV^{-1}\,cm^{-2}\,s^{-1}]$}  & \\
\hline
LHAASO J0534+2202 & Crab Nebula     &2.0$\rm ^{a}$  &0.13 &PS &3.7  &--- &--- &0.76 & Ten-year$\rm ^{d}$\\
                  &                 &           &0.13 &PS &4.74  &--- &--- &--- & Eight-year$\rm ^{e}$\\
                  &                 &           &0.13 &PS &10.7 &11.8 &10.7 & --- & TeV PWNe$\rm ^{g}$ \\
                  &                 &           &0.10 &PS &---  &--- &117 &--- & SNR with PWN$\rm ^{h}$\\
\hline
LHAASO J1825-1326 & HESS J1826-130  &4.0$\rm^{a}$ &0.44 &PS    &13  &--- &25 &--- & $\rm ANTARES\,\&\,IceCube^{f}$\\
                  & HESS J1825-137  &3.9$\rm^{a}$ &0.39 &0.461 &10.7 &11.8 &10.7 &--- & TeV PWNe\\
                  & 1HWC J1825-133  &3.9$\rm^{a}$ &0.21 &0.5   &---  &--- &0.948 &--- & HAWC Hotspots$\rm ^{h}$ \\
\hline
LHAASO J1839-0545 & HESS J1841-055  &--- &0.52 &PS  &4.8  &--- &--- &3.28 & Ten-year\\
                  & 1HWC J1838-060  &--- &0.29 &0.5 &---  &--- &0.948 &--- & HAWC Hotspots\\
\hline
LHAASO J1843-0338 & HESS J1843-033  &--- &0.35 &PS  &2.5  &--- &--- &1.09 & Ten-year\\
                  & 1HWC J1844-031c &--- &0.60 &0.5 &---  &--- &0.948 &--- & HAWC Hotspots \\
                  & 1HWC J1842-046c &--- &0.98 &0.5 &---  &--- &0.948 &--- & HAWC Hotspots \\
\hline
LHAASO J1849-0003 & HESS J1849-000  &7$\rm ^{a}$ &0.09 &PS   &2.2  &--- &--- &1.01 & Ten-year\\ 
                  & IGR J18490-0000 &7$\rm ^{a}$ &0.11 &0.09 &10.7  &11.8 &10.7 &--- & TeV PWNe\\
\hline
LHAASO J1908+0621 & MGRO J1908+06   &---        &0.21 &PS  &5.7  &--- &--- &2.11 & Ten-year\\
                  &                 &           &0.10 &PS  &7.62  &--- &--- &--- & Eight-year\\
                  &                 &           &0.49 &1.3 &---  &--- &3.98 &--- & Milagro Six$\rm ^{h}$\\
                  & 1HWC J1907+062c &---        &0.29 &0.5 &---  &--- &0.948 &--- & HAWC Hotspots\\
\hline
LHAASO J2018+3651 & MGRO J2019+37   &--- &0.09 &PS   &4.0  &--- &--- &0.69 & Ten-year\\
                  &                 &    &0.38 &PS   &4.54  &--- &--- &--- & Eight-year\\
                  &                 &    &0.08 &0.75 &10.7 &11.8 &10.7 &--- & TeV PWNe\\
                  &                 &    &0.16 &0.64 &---  &--- &3.98 &--- & Milagro Six\\
\hline
LHAASO J2032+4102 & 2HWC J2031+415  &1.8$\rm ^{a}$ &0.47 &PS    &9.2  &--- &--- &1.42 & Ten-year\\ 
                  & Cyg OB2         &1.5$\rm ^{b}$ &0.18 &PS    &7.64  &--- &--- &--- & Eight-year\\
                  & Cyg X-3         &7.4$\rm ^{c}$ &0.10 &PS    &8.20  &--- &--- &--- & Eight-year\\
                  & MGRO J2031+41   &---           &0.39 &1.5   &---  &--- &3.98 &--- & Milagro Six\\
                  & TeV J2032+4130  &1.8$\rm ^{a}$ &0.46 &0.158 &10.7  &11.8 &10.7 &--- & TeV PWNe\\
\hline
LHAASO J2226+6057 &Boomerang        &0.8$\rm ^{a}$ &0.30 &0.22  &10.7  &11.8 &10.7 &--- & TeV PWNe\\
\specialrule{0em}{1pt}{1pt}
\hline
\end{tabular*}
\caption{Neutrino source candidates within one degree away from the measured center of each LHAASO source. Column 3 shows the distances between the candidates and the Earth, which are derived from $\rm ^{a}$ \href{http://tevcat2.uchicago.edu}{TeVCat}; $\rm ^{b}$ \cite{2009MNRAS.400..518M}; $\rm ^{c}$ \cite{2016ApJ...830L..36M}. Column 4 shows the angular distances between the candidates and LHAASO sources. Column 5 shows the extensions of neutrino source candidates considered in the neutrino source searches. Column 6-9 show the 90\% C.L. upper limits of neutrino flux parameterized as: $dN_{\nu_{\mu}+\bar{\nu}_{\mu}}/dE_{\nu}=\Phi_{0,\gamma}^{90\%}\cdot(E_{\nu}/100\,{\rm TeV})^{-\gamma}\times10^{-17}\,{\rm TeV^{-1}~cm^{-2}~s^{-1}}$. The coordinates, extensions, and neutrino flux of neutrino source candidates are derived from SLSes: $\rm ^{d}$ the ten-year point source search \citep{2020PhRvL.124e1103A}, $\rm ^{e}$ the eight-year point source search \citep{2019EPJC...79..234A},  $\rm ^{f}$ the ANTARES \& IceCube combined search \citep{2020ApJ...892...92A} and SSSes: $\rm ^{g}$ the TeV PWNe search \citep{2020ApJ...898..117A}, $\rm ^{h}$ the Galactic catalog searches \citep{2017ApJ...849...67A}. The upper limits for $E^{-3}$ spectrum in the ten-year search are taken from Figure 3 of \cite{2020PhRvL.124e1103A}. The data point of specific source in Figure 3 can be identified with source declination and local pre-trial p-value in the Table III of \cite{2020PhRvL.124e1103A}. The lower p-value indicates the higher upper limit for the sources with similar declination angles. As for TeV PWNe search, only the upper limits for the inverse age weighting scheme are displayed.}\label{table01}
\end{center}
\end{sidewaystable}


\clearpage

\appendix

\section{Weighting Schemes}\label{appendix:A}



 The TeV PWNe search gave the upper limits of stacking neutrino flux from 35 identified TeV PWNe under four weighting schemes \citep{2020ApJ...898..117A}.  Neutrino flux from each source in the catalog is assumed to be the same in the equal weighting, proportional to gamma-ray flux at $\rm 1\,TeV$ in the gamma-ray flux weighting, proportional to inverse age of pulsar in the inverse age weighting and proportional to pulsar frequency in the frequency weighting. 


The Galactic catalog searches gave the upper limit of stacking neutrino flux over five Galactic catalogs respectively \citep{2017ApJ...849...67A}. In the Milagro catalog, neutrino flux from each source is assumed to follow the model of \cite{2009NIMPA.602..117K}. In the HAWC catalog, neutrino flux from each source is assumed to be proportional to the gamma-ray flux at pivot energy. In the other three SNR catalogs, neutrino flux from each source is assumed to be equal. 


\section{TeV Gamma-ray Observations}\label{appendix:B}

\paragraph{LHAASO J1825-1326} We take the gamma-ray spectrum of eHWC J1825-134 instead of that of 1HWC J1825-133, because 1HWC J1825-133 is reported as 2HWC J1825-134 in the 2HWC catalog \citep{2017ApJ...843...40A}, while 2HWC J1825-134 is only 0.07 degree from eHWC J1825-134 ($\rm >56\,TeV$) \citep{2020PhRvL.124b1102A}. The flux data points of HESS J1826-130 are raised by a factor of 2.37 from the original values, if we consider a 2-d Gaussian intrinsic extension of $0.21^{\circ}$ with the integration region of $0.22^{\circ}$ \citep{2020A&A...644A.112H}.

\paragraph{LHAASO J1839-0545} We only take the spectrum of HESS J1841-055, because 1HWC J1838-060 overlaps with the extension of HESS J1841-055 and its differential flux normalization is compatible with the previous measurements of HESS J1841-055 \citep{2016ApJ...817....3A}. The flux data points of HESS J1841-055 are raised by a factor of 1.14 based on its intrinsic extension of $0.41^{\circ}$ ($0.25^{\circ}$) along the major (minor) axis and the integration region of $0.7^{\circ}$ \citep{2008A&A...477..353A}.

\paragraph{LHAASO J1843-0338} We take the observation of 2HWC J1844-032, because 1HWC J1844-031c is reported as 2HWC J1844-032 in the 2HWC gamma-ray catalog \citep{2017ApJ...843...40A}.

\paragraph{LHAASO J1849-0003} We only take the spectrum of HESS J1849-000, because HESS J1849-000 is in spatial coincidence with IGR J18490-0000 \citep{2008AIPC.1085..312T}.

\paragraph{LHAASO J1908+0621} We take the spectrum of eHWC J1907+063. 1HWC J1907+062c is reported as 2HWC J1908+063 \citep{2017ApJ...843...40A}, while 2HWC J1908+063 is a point source only 0.17 degree from eHWC J1907+063 ($>$56 TeV) \citep{2020PhRvL.124b1102A}. The flux data points of HESS J1908+063 are raised by a factor of 1.51 for its intrinsic extension of $0.34^{\circ}$ and the integration region of $1.3^{\circ}$\citep{2009A&A...499..723A}.

\paragraph{LHAASO J2018+3651} We take the spectrum of VER J2019+368 because it constitutes the bulk of the emission of MGRO J2019+37 \citep{2018ApJ...861..134A}. The flux data points of VER J2019+368 is raised by a factor of 2.53 based on its intrinsic extension of $0.34^{\circ}$ ($0.14^{\circ}$) along the major (minor) axis and the integration region of $0.23^{\circ}$. We also use the observation of eHWC J2019+368 because its angular distance to LHAASO J2018+3651 is only 0.17 degree and it also has 100 TeV measurements \cite{2020PhRvL.124b1102A}.

\paragraph{LHAASO J2032+4102} The energy spectrum of TeV J2032+4130 is raised by different factors of 1.79, 4.64 and 2.72 for the flux data points below $\rm 1.46\,TeV$, between $\rm 1.46\,TeV$ and $\rm 2.76\,TeV$, and above $\rm 2.76\,TeV$ respectively, according to its intrinsic extensions at different energy bands \citep{2014ApJ...783...16A}.

\paragraph{LHAASO J2226+6057} The VERITAS flux data points of G106.3+2.7 are raised by a factor of 1.59 based on its intrinsic extension of $0.27^{\circ}$ ($0.18^{\circ}$) along the major (minor) axis with an integration region of $0.32^{\circ}$ \citep{2009ApJ...703L...6A}.

For the NSCs not mentioned above, we simply take their gamma-ray measurements without adjustments.

\section{Gamma-ray Flux from TeV PWNe}\label{appendix:C}

There are 6 PWNe associated with LHAASO sources. We fit the gamma-ray spectra of these sources from TeV to sub-PeV. The best spectral fits are selected among the single power law (PL) model $dN/dE\propto(E/10\,\rm{TeV})^{-\gamma_{1}}$, the exponential cutoff power law (ECPL) model $dN/dE\propto(E/10\,\rm{TeV})^{-\gamma_{1}}e^{-E/E_{c}}$ and the log-parabola (LOGP) model $dN/dE\propto(E/10\,\rm{TeV})^{-\gamma_{1}-\gamma_{2}\,\rm{log}(E/10\,\rm{TeV})}$ based on the corrected Akaike Information Criterion ($\rm AIC_{C}$) \citep{10.1093/biomet/78.3.499}. Both the gamma-ray measurements of the identified PWNe and the associating LHAASO sources are used in the fitting except for HESS J1825-137. The observations by LHAASO are not considered for HESS J1825-137, because HESS J1825-137 is just one of three sources composing the gamma-ray hotspot and the gamma-ray flux at 100 TeV is not dominated by this PWN \citep{2021ApJ...907L..30A}. The best-fit parameters are shown in \autoref{table02} (row 1-6).


There are 7 PWNe in the data taken region $-15^{\circ}<\delta<75^{\circ}$ ($\delta$ is declination angle) but not discovered by KM2A half-array. The upper limit on the gamma-ray flux is $1.1\times10^{-13}\,\rm{TeV\,cm^{-2}\,s^{-1}}$ between 100 TeV and 178 TeV \citep{2021Natur.594...33C}. We fit the upper limit of energy flux at the 100-178 TeV bin with the ECPL model, following the spectral index used in the TeV PWNe search \citep{2020ApJ...898..117A}. The best-fit cutoff energies are also the upper limits of cutoff energies $E_{c}$. The best-fit $E_{c}$ is negative for SNR G054.1+00.3, so we just follow its spectrum in the TeV PWNe search. The upper limits of $E_{c}$ for the other 6 sources are shown in \autoref{table02} (row 7-13).


The correction for the interstellar absorption is necessary for the 12 PWNe in \autoref{table02}. The distances to 8 PWNe are available from \href{http://tevcat2.uchicago.edu}{TeVCat}. As for the other 4 PWNe, we place them at the border of the ISRF model (a cylinder with radius 24 kpc and half-height 10 kpc) given by \citet{2017MNRAS.470.2539P} to get the maximum absorption to gamma-rays around 100 TeV. So we can get a conservative upper limit on gamma-ray flux from these PWNe before absorption. The formula to calculate gamma-ray opacity follows the equations in section 3 of \cite{2006ApJ...640L.155M}. This upper limit is also hold when the gamma-rays from LHAASO sources are not from the PWNe, but the SNRs or YMCs associated \citep[e.g.][]{2019NatAs...3..561A, 2020arXiv201211531G}. 

As for the other 22 PWNe outside the data taken region, we simply use the same spectra in the TeV PWNe search \citep{2020ApJ...898..117A}. The sum of gamma-ray spectra obtained above is shown as the black solid line in \autoref{figure_pwn}.

\begin{center}
\begin{table}[htbp]
\begin{center}
\begin{tabular}{l c c c c c c c}
\hline
\hline
PWN & $\gamma_{1}$ & $\gamma_{2}$ & $E_{\rm c}$ & Model & Distance\\
    &              &              & [TeV]   &       & [kpc]\\
\hline
Crab Nebula & 2.86 & 0.20 & --- & LOGP& 2.0\\
HESS J1825-137 & 2.42 & --- & 31 & ECPL& 3.9\\
IGR J18490-0000 & 1.99 & --- & --- & PL & 7.0\\
MGRO J2019+37 & 2.2 & 0.83 & --- & LOGP & 24.8$^{\rm*}$\\
TeV J2032+4130 & 2.36 & 0.41 & --- & LOGP & 1.8\\
Boomerang & 2.29 & 0.35 & --- & LOGP & 0.8\\
CTA1      & 2.2 & --- & 214 & ECPL & 1.4\\
Geminga   & 2.23 & --- & 74  & ECPL & 0.25 \\
2HWC J0700+143 & 2.17 & --- & 196  & ECPL & 17.2$^{\rm*}$\\
HESS J1831-098 & 2.1 & --- & 188 & ECPL & 30.9$^{\rm*}$\\
HESS J1837-069 & 2.54 & --- & 80  & ECPL & 6.6 \\
MAGIC J1857.2+0263 & 2.2 & --- & 119  & ECPL & 29.7$^{\rm*}$\\
\specialrule{0em}{1pt}{1pt}
\hline
\end{tabular}
\caption{The gamma-ray spectral parameters of TeV PWNe constrained by LHAASO's observation. Row 1-6: The best-fit spectral parameters for the PWNe associated with LHAASO sources. Row 7-12: The upper limits of cutoff energy $E_{\rm c}$ for the PWNe in the data taken region but not discovered by LHAASO-KM2A. These parameters are corrected for the interstellar absorption. $^{*}$the distance to the source at the border of ISRF given by \citet{2017MNRAS.470.2539P}.}\label{table02}
\end{center}
\end{table}
\end{center}

\end{document}